\documentclass[dvips,12pt,a4paper]{article}
\usepackage{a4wide}
\usepackage{epsfig}
\usepackage{amsmath}
\usepackage{latexsym}

  \newlength{\absize}
\newcommand{\dd}{\mbox{{\rm d}}}

\newcommand{\Lumint}{{\cal L}_{\rm int}}

\allowdisplaybreaks 

\catcode`@=11
\def\citer{\@ifnextchar [{\@tempswatrue\@citexr}{\@tempswafalse\@citexr[]}}

\def\@citexr[#1]#2{\if@filesw\immediate\write\@auxout{\string\citation{#2}}\fi
  \def\@citea{}\@cite{\@for\@citeb:=#2\do
    {\@citea\def\@citea{--\penalty\@m}\@ifundefined
       {b@\@citeb}{{\bf ?}\@warning
       {Citation `\@citeb' on page \thepage \space undefined}}%
\hbox{\csname b@\@citeb\endcsname}}}{#1}}
\catcode`@=12

\begin{document}
  \thispagestyle{empty}
  \pagestyle{empty}
  \renewcommand{\thefootnote}{\fnsymbol{footnote}}
\newpage\normalsize
    \pagestyle{plain}
    \setlength{\baselineskip}{4ex}\par
    \setcounter{footnote}{0}
    \renewcommand{\thefootnote}{\arabic{footnote}}
\newcommand{\preprint}[1]{%
  \begin{flushright}
    \setlength{\baselineskip}{3ex} #1
  \end{flushright}}
\renewcommand{\title}[1]{%
  \begin{center}
    \LARGE #1
  \end{center}\par}
\renewcommand{\author}[1]{%
  \vspace{2ex}
  {\Large
   \begin{center}
     \setlength{\baselineskip}{3ex} #1 \par
   \end{center}}}
\renewcommand{\thanks}[1]{\footnote{#1}}
\renewcommand{\abstract}[1]{%
  \vspace{2ex}
  \normalsize
  \begin{center}
    \centerline{\bf Abstract}\par
    \vspace{2ex}
    \parbox{\absize}{#1\setlength{\baselineskip}{2.5ex}\par}
  \end{center}}

\begin{flushright}
{\setlength{\baselineskip}{2ex}\par

}
\end{flushright}
\vspace*{4mm}
\vfill
\title{Model independent constraints on\\ 
contact interactions from LEP2\thanks{Partially supported by MURST (Italian 
Ministry of University, Scientific Research and Technology).}}

\vfill
\author{
A.A. Pankov$^{a,b,c}$ {\rm and}
N. Paver$^{c,}$\footnote{E-mail address: nello.paver@ts.infn.it}}
\begin{center}
$^a$ Pavel Sukhoi Technical University, 
     Gomel 246746, Belarus \\
$^b$ International Centre for Theoretical Physics, 34100 Trieste, Italy   \\
$^c$ Dipartimento di Fisica Teorica, Universit\`a di Trieste and \\
Istituto Nazionale di Fisica Nucleare, Sezione di Trieste, 34100 
Trieste, Italy
\end{center}
\vfill
\abstract
{We quantitatively discuss the possibility of deriving model-independent
constraints on the general four-fermion contact interaction couplings, from
the currently available data on the two-fermion production processes
$e^+e^-\to\mu^+\mu^-$, $b\bar{b}$ and $c\bar{c}$ with
unpolarized initial beams. The method is essentially based on particular,
simple, combinations of the measured total cross section and forward-backward
asymmetry that allow partial separation of the helicity cross sections, and the
combination of experimental data obtained at the different energies of TRISTAN,
LEP1 and LEP2.}

\vspace*{20mm}
\setcounter{footnote}{0}
\vfill

\newpage
    \setcounter{footnote}{0}
    \renewcommand{\thefootnote}{\arabic{footnote}}
    \setcounter{page}{1}
High precision data on fermion-pair production by $e^+e^-$ collisions at
LEP is regarded both as a powerful test of the Standard Model (SM)
and as an interesting tool to severely constrain the parameters of
non-standard dynamics that might manifest themselves through deviations of the
measured observables from the SM predictions \cite{lep}.
\par
In particular, this is the case of the $SU(3)\times SU(2)\times U(1)$
symmetric $eeff$ contact-interaction Lagrangian with helicity-conserving
and flavor-diagonal fermion currents that can be expressed as \cite{eichten}:
\begin{equation}
{\cal L}=\sum_{\alpha\beta}{g^2_{\rm eff}}\hskip2pt
\epsilon_{\alpha\beta}\left(\bar e_{\alpha}\gamma_\mu e_{\alpha}\right)
\left(\bar f_{\beta}\gamma^\mu f_{\beta}\right),
\label{lagra}
\end{equation}
where generation and color indices are not explicitly indicated,
$\alpha,\beta={\rm L,R}$ denote left- or right-handed fermion helicities, and
the parameters $\epsilon_{\alpha\beta}=\pm 1/\Lambda_{\alpha\beta}^2$ specify
the chiral structure of the individual interactions, with
$\Lambda_{\alpha\beta}$ some high energy scales that determine the size of the
effects. Conventionally, the scales of $\Lambda$'s are chosen by conventionally
fixing $g^2_{\rm eff}/4\pi=1$ as a
reminder that this new interaction, originally proposed for compositeness,
would become strong at the reaction energy
$\sqrt s\sim\Lambda_{\alpha\beta}$. In fact, more
generally, Eq.~(\ref{lagra}) can be considered as an effective Lagrangian that
parametrizes the effects at the `low' energy $\sqrt s$ of any
non-standard dynamics acting at the much larger scales
$\Lambda_{\alpha\beta}\gg \sqrt s$,
at the leading order in $\sqrt s/\Lambda_{\alpha\beta}$. In addition to the
remnant compositeness binding force mentioned above, familiar examples are the
exchanges of a heavy $Z^\prime$ \cite{barger} and of a heavy
leptoquark \cite{altarelli}. From this point of view, the scales
$\Lambda_{\alpha\beta}$
define the standard to compare the sensitivity of measurements to the various
kinds of new interactions.
\par
Referring to the processes under consideration($f\ne e$, $t$):
\begin{equation}
e^++e^-\to f+\bar{f}, \label{proc} \end{equation}
once combined with the SM $\gamma$- and $Z$-exchanges, the contact Lagrangian
${\cal L}$ should ``indirectly'' manifest itself by modifications of observables
from the SM predictions and, clearly, the numerical comparison of such
deviations to the experimental accuracies quantitatively determines the
attainable reach in the free mass scales $\Lambda_{\alpha\beta}$
or, equivalently, the experimental sensitivity to the new coupling constants
$\epsilon_{\alpha\beta}$.
\par
In practice, the situation is complicated by the fact that, for a given flavor
$f$, Eq.~(\ref{lagra}) defines eight individual,
independent, models corresponding to the combinations of the four chiralities
$\alpha,\beta$ with the $\pm$ signs of the $\epsilon$'s, and the general
contact interaction could be any linear combination of these models.
Accordingly, the aforementioned deviations of from the SM predictions
simultaneously depend on all four-fermion effective couplings and,
for a fixed value of the energy $\sqrt s$, their straightforward comparison to
the experimental uncertainties {\it a priori} could only produce numerical
correlations among the possible values of the different couplings, rather than
separate, and restricted, allowed regions around the SM limit
$\epsilon_{\alpha\beta}=0$. This could be obtained only by a procedure based on
suitable observables and/or the analysis of appropriate samples of experimental
data.
\par
Indeed, the simplest, and commonly adopted, procedure consists in assuming
non-zero values for just one of the $\epsilon_{\alpha\beta}$ at a time, and in 
constraining it to a finite interval by essentially a $\chi^2$ fit analysis
of the total cross section and the forward-backward asymmetry, while all the
remaining parameters are set equal to zero \cite{barger,kroha}. In this way, 
only tests of the aforementioned particular models can be performed.
\par 
On the other hand, as emphasized above, it is desirable to perform a general,
and model-independent, kind of analysis of the experimental data that
simultaneously includes all terms of Eq.~(\ref{lagra}) as free parameters and,
at the same time, allows to disentangle their contributions to the basic
observables to the largest possible extent in order to derive separate
constraints. In particular, this procedure would avoid potential cancellations
between different contributions that might considerably weaken the numerical
constraints. The ideal solution to this problem would be represented by the
longitudinal initial electron beam polarization, that would enable us to
experimentally extract the individual helicity amplitudes $A_{\alpha\beta}$
of process (\ref{proc}), that by definition are directly related to the single
$eeff$ contact coupling $\epsilon_{\alpha\beta}$ and, therefore, depend on the
minimal set of free independent parameters \cite{pankov}. Unfortunately, such
a procedure cannot be applied to the LEP data, that refer to measurements of
the total cross section $\sigma$ and of the forward-backward asymmetry
$A_{\rm FB}$ with unpolarized electron and positron beams.
\par 
In this case, we adopt an approach based on two particular observables,
$\sigma_+$ and $\sigma_-$, that are simple
combinations of the ``conventional'' ones $\sigma$ and $A_{\rm FB}$, and
allow to partially disentangle
the contributions of the terms with different chiralities in
Eq.~(\ref{lagra}). Since the separation is only partial, by themselves
$\sigma_+$ and $\sigma_-$ would still lead to correlations among pairs of
contact interaction couplings squared rather than a well-defined, and
restricted, allowed region. This restricted area could be derived by a global
analysis that supplements the recent LEP2 data on $\sigma$ and $A_{\rm FB}$
with the measurements of the same observables at the quite different energies
of LEP1 and of TRISTAN, taking advantage of the expected energy-dependence of
the deviations from the SM due to the new interaction (\ref{lagra}),
entirely determined by well-known SM parameters.
\par
Limiting ourselves to the cases $f\neq e, t$ and neglecting all fermion masses
with respect to $\sqrt s$, the amplitude for process (\ref{proc}) is determined
by the Born $\gamma$ and $Z$ exchanges in the $s$ channel plus the
contact-interaction term of Eq.~(\ref{lagra}). With $\theta$ the angle between
the incoming electron and the outgoing fermion in the c.m. frame, the
differential cross section reads \cite{zeppenfeld2}:
\begin{equation}
\frac{\dd\sigma}{\dd\cos\theta}
=\frac{3}{8}
\left[(1+\cos\theta)^2 {\sigma}_+
+(1-\cos\theta)^2 {\sigma}_-\right].
\label{cross}
\end{equation}
In terms of helicity cross sections $\sigma_{\alpha\beta}$ (with
$\alpha,\beta={\rm L,R}$):
\begin{eqnarray}
{\sigma}_{+}&=&\frac{1}{4}\,
\left(\sigma_{\rm LL}+\sigma_{\rm RR}\right),
\label{s+} \\
{\sigma}_{-}&=&\frac{1}{4}\,
\left(\sigma_{\rm LR}+\sigma_{\rm RL}\right).
\label{s-}
\end{eqnarray}
In Eqs.~(\ref{s+}) and (\ref{s-}):
\begin{equation}
\sigma_{\alpha\beta}=N_C\sigma_{\rm pt}
\vert A_{\alpha\beta}\vert^2,
\label{helcross}
\end{equation}
where $N_C\simeq 3(1+\alpha_s/\pi)$ for quarks and $N_C=1$ for leptons,
respectively, and $\sigma_{\rm pt}\equiv\sigma(e^+e^-\to\gamma^\ast\to l^+l^-)
=4\pi\alpha_{e.m.}^2/3s$. The helicity amplitudes $A_{\alpha\beta}$ can be
written as
\begin{equation}
A_{\alpha\beta}=Q_e Q_f+g_\alpha^e\,g_\beta^f\,\chi_Z+
\frac{s}{\alpha_{e.m.}}\epsilon_{\alpha\beta},
\label{amplit}
\end{equation}
where: $\chi_Z=s/(s-M^2_Z+iM_Z\Gamma_Z)$ is $Z$ propagator;
$g_{\rm L}^f=(I_{3L}^f-Q_f s_W^2)/s_W c_W$ and
$g_{\rm R}^f=-Q_f s_W^2/s_W c_W$
are the SM left- and right-handed fermion couplings of the $Z$
with $s_W^2=1-c_W^2\equiv \sin^2\theta_W$; $Q_f$ are the fermion electric
charges.
\par
The ``conventional'' observables $\sigma$ and $A_{\rm FB}$ are given by the
relations:
\begin{equation}
\label{sigma}
\sigma
={\sigma}_{+}+{\sigma}_{-}=\sigma_{\rm F}+\sigma_{\rm B}
=\frac{1}{4}\left(\sigma_{\rm LL}+\sigma_{\rm RR}
+\sigma_{\rm LR}+\sigma_{\rm RL}\right);
\end{equation}
and
\begin{equation}\sigma_{\rm FB}\equiv\sigma\, A_{\rm FB}
=\sigma_{\rm F}-\sigma_{\rm B}
=\frac{3}{4}\left({\sigma}_{+}-{\sigma}_{-}\right) 
=\frac{3}{16}\left(\sigma_{\rm LL}+\sigma_{\rm RR}
-\sigma_{\rm LR}-\sigma_{\rm RL}\right).
\label{sigmafb}
\end{equation}
Finally, the relation to $\sigma_{\pm}$ is:
\begin{equation}
\sigma_{\pm}=
\frac{\sigma}{2}\hskip 2pt \left(1\pm\frac{4}{3}\hskip 1pt A_{\rm FB}\right).
\label{sig+-}
\end{equation}
Taking Eq.~(\ref{amplit}) into account, Eqs.~(\ref{sigma}) and (\ref{sigmafb})
show that the ``conventional'' observables simultaneously depend on all four
contact interaction couplings and consequently do not allow a
model-independent analysis, but only the simplified one-parameter fit of
individual models mentioned above. Instead, as shown by Eqs.~(\ref{s+}) and
(\ref{s-}), the situation is definitely improved by considering the
combinations $\sigma_+$ and $\sigma_-$, each one depending on just pairs of
contact interaction parameters by construction. Consequently, such pairs of
coupling constants can be separately constrained and furthermore, the
combination of data on, respectively, $\sigma_+$ and $\sigma_-$ at different
energies will allow to further restrict such separate bounds in a
model-independent way.
\par
To this purpose, one has to quantitatively assess the sensitivity to the
contact interaction couplings of $\sigma_+$ and $\sigma_-$, that ultimately
will specify the ``significance'', i.e., the attainable reach on these
parameters for given experimental uncertainty on the basic observables. With
${{\cal O}=\sigma_{\pm}}$, such sensitivity can be defined as
\begin{equation}
{\cal S}=\frac{\vert\Delta{\cal O}\vert}{\delta{\cal O}},
\label{signif}
\end{equation}
where $\Delta{\cal O}={\cal O}^{SM+CI}-{\cal O}^{SM}$ represents the deviation
from the SM prediction induced by the Lagrangian of Eq.~(\ref{lagra}) and
$\delta{\cal O}$ is the corresponding experimental uncertainty, combining
statistical and systematical uncertainties. In our application,
we express $\sigma_{\pm}^{SM}$ in terms of improved Born
amplitudes \cite{hollik,altarelli2}, such that the form of the
previous formulae remains the same, with the values $m_{\rm top}=175$~GeV and
$m_H=100$~GeV. Regarding the experimental input, the values of $\sigma_+$ and
$\sigma_-$ as well as their uncertainties $\delta\sigma_+$ and
$\delta\sigma_-$ are reconstructed 
from the measured $\sigma$ and $A_{\rm FB}$
at the different energies {\it via} Eq.~(\ref{sig+-}). It can be seen that, to
a very good approximation, the correlation between the uncertainties on
$\sigma$ and $A_{\rm FB}$ is negligible. In all numerical analyses, we take
initial- and final-state radiation into account by the programs
ZFITTER and ZEFIT \cite{zfitter} properly adapted to the present
case of contact interactions, with the experimentally used cuts on the initially
radiated photon energy.
\begin{table} [h]
\centering
\caption{Integrated luminosity per experiment collected during the runs 
at TRISTAN, LEP1 and LEP2.}
\medskip
{\small
\begin{tabular}{|c|c|c|}
\hline
collider & $E_{CM}$ (GeV) &  $\Lumint$ $[{\rm pb}^{-1}]$ \\
\hline\hline
Tristan \cite{tristan} & 58 &  300 \\
\hline
\cline{2-3}
LEP1 \cite{lep1}& peak-3  & 1 \\ \cline{2-3}
$Z$\ {\rm line shape}    & peak-2 & 21\\ \cline{2-3}
                       & peak-1  & 1 \\ \cline{2-3}
                       & peak+1  & 1 \\ \cline{2-3}
                       & peak+2 & 21\\ \cline{2-3}
                       & peak+3  & 1 \\ \hline
LEP2 \cite{lep2}& 130   & 3 \\ \cline{2-3}
                       & 136   & 3 \\ \cline{2-3}
                       & 161   & 10\\ \cline{2-3}                       
                       & 172   & 10\\ \cline{2-3}
                       & 183   & 53\\ \cline{2-3}
                       & 189   & 158\\ \cline{2-3}
                       & 192   & 25\\ \cline{2-3}
                       & 196   & 76\\ \cline{2-3}
                       & 200   & 83\\ \cline{2-3}
                       & 202  & 40 \\ \hline
\end{tabular} } 
\end{table}
\par
In Table 1, we show the energy points
considered in the subsequent numerical analysis and the corresponding collected
luminosities that determine the statistical uncertainties (in many cases 
dominant), referring for definiteness to the experiments DELPHI at LEP and 
VENUS and AMY at TRISTAN. As anticipated, our derivation of model-independent
bounds on the contact
interaction couplings will make use of the combinations of constraints at
different energies. In Figs. 1-3, we show as a representative case the values
of $\cal S$ defined in Eq.~(\ref{signif}) for given reference values
$\epsilon_{\alpha\beta}=5\ 10^{-2}$, at the energy points listed in
Table 1, and using the corresponding experimental uncertainties
$\delta\sigma_{\pm}$.

\begin{figure}[htb]
\refstepcounter{figure}
\label{Fig1}
\addtocounter{figure}{-1}
\begin{center}
\setlength{\unitlength}{1cm}
\begin{picture}(12,6.5)
\put(-3.,0.0)
{\mbox{\epsfysize=8.5cm\epsffile{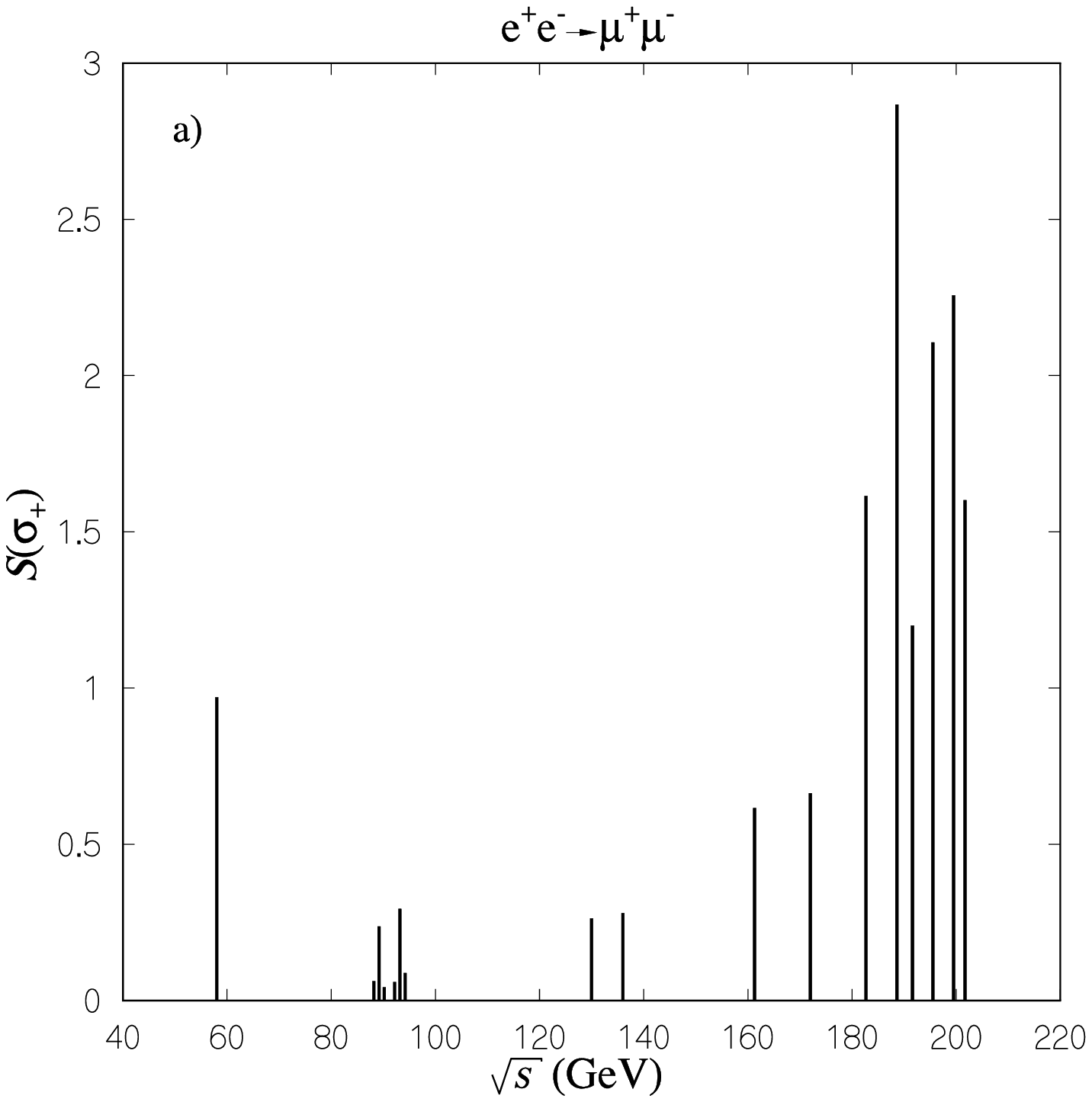}}
 \mbox{\epsfysize=8.5cm\epsffile{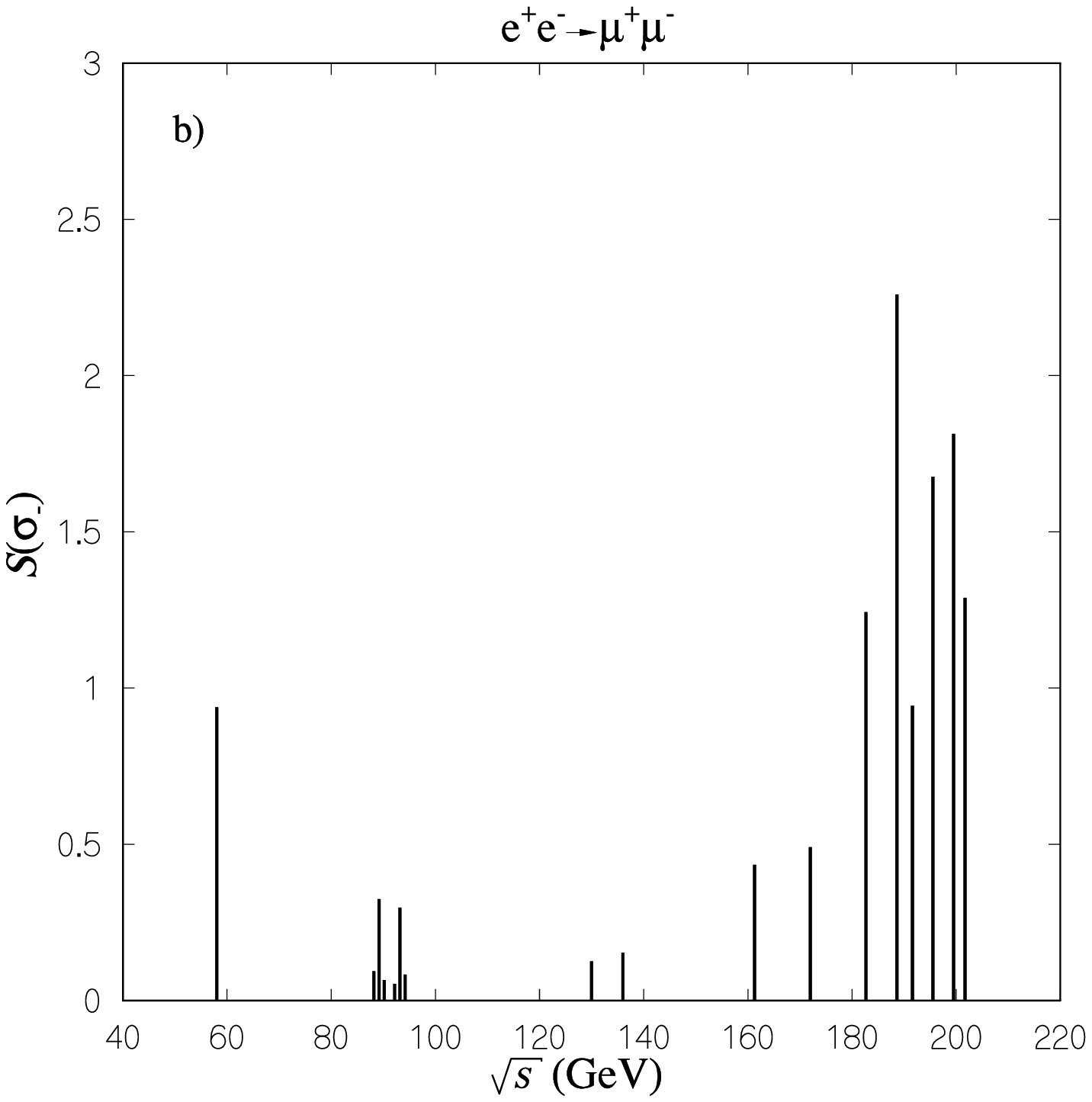}}}
\end{picture}
\caption{Sensitivity of the observables $\sigma_\pm$ to contact interaction
parameters $\epsilon_{\rm RR}=5\ 10^{-2}$, $\epsilon_{\rm LL}=0$ (a) and
$\epsilon_{\rm RL}=5\ 10^{-2}$, $\epsilon_{\rm LR}=0$
(b) for the process $e^+e^-\to\mu^+\mu^-$ at the energy points listed in
Table 1, and using the experimental uncertainties in
Refs.~\cite{tristan,lep1,lep2}}.
\end{center}
\end{figure}
\begin{figure}[htb]
\refstepcounter{figure}
\label{Fig2}
\addtocounter{figure}{-1}
\begin{center}
\setlength{\unitlength}{1cm}
\begin{picture}(12,7.5)
\put(-3.,0.0)
{\mbox{\epsfysize=8.5cm\epsffile{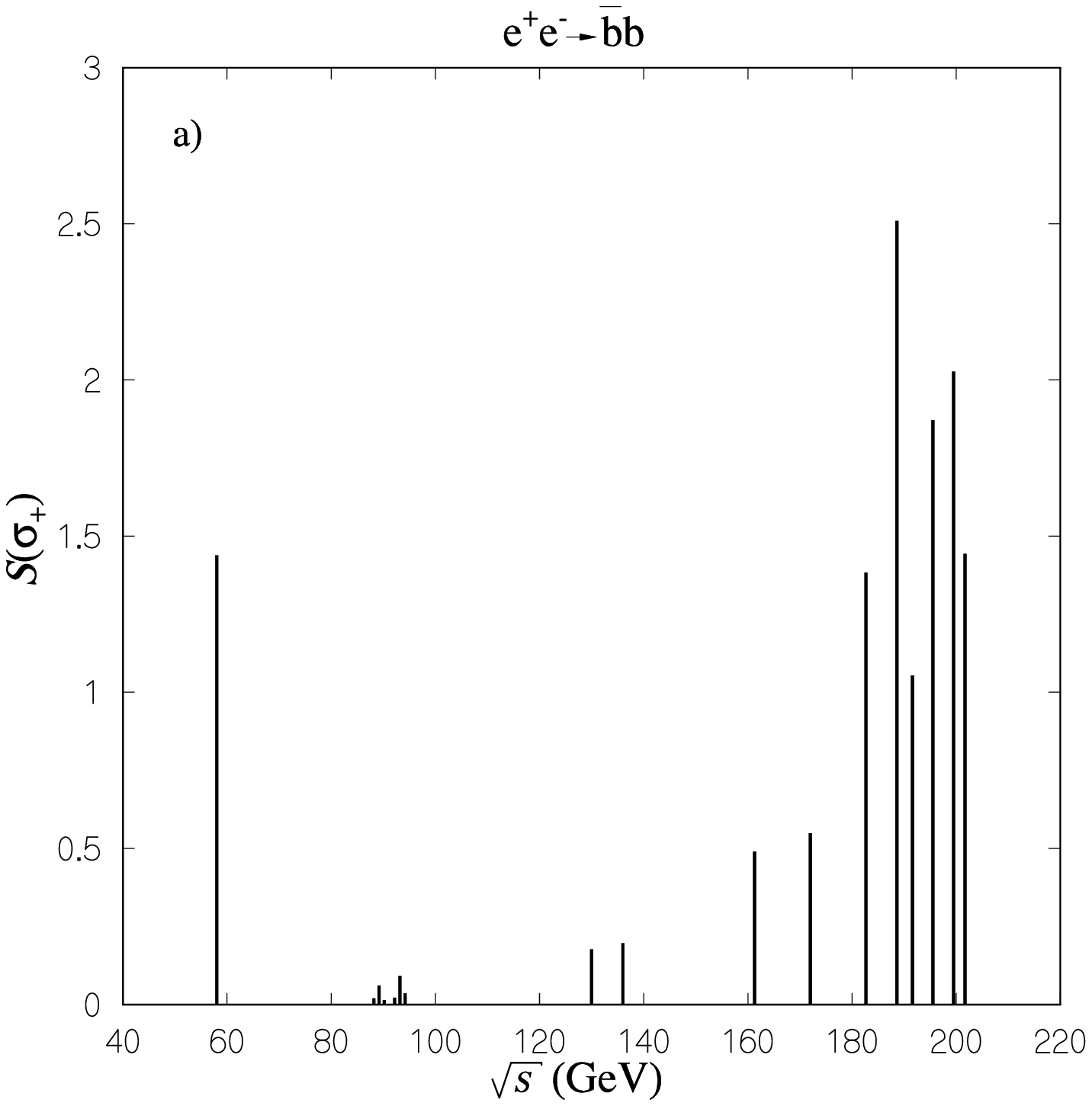}}
 \mbox{\epsfysize=8.5cm\epsffile{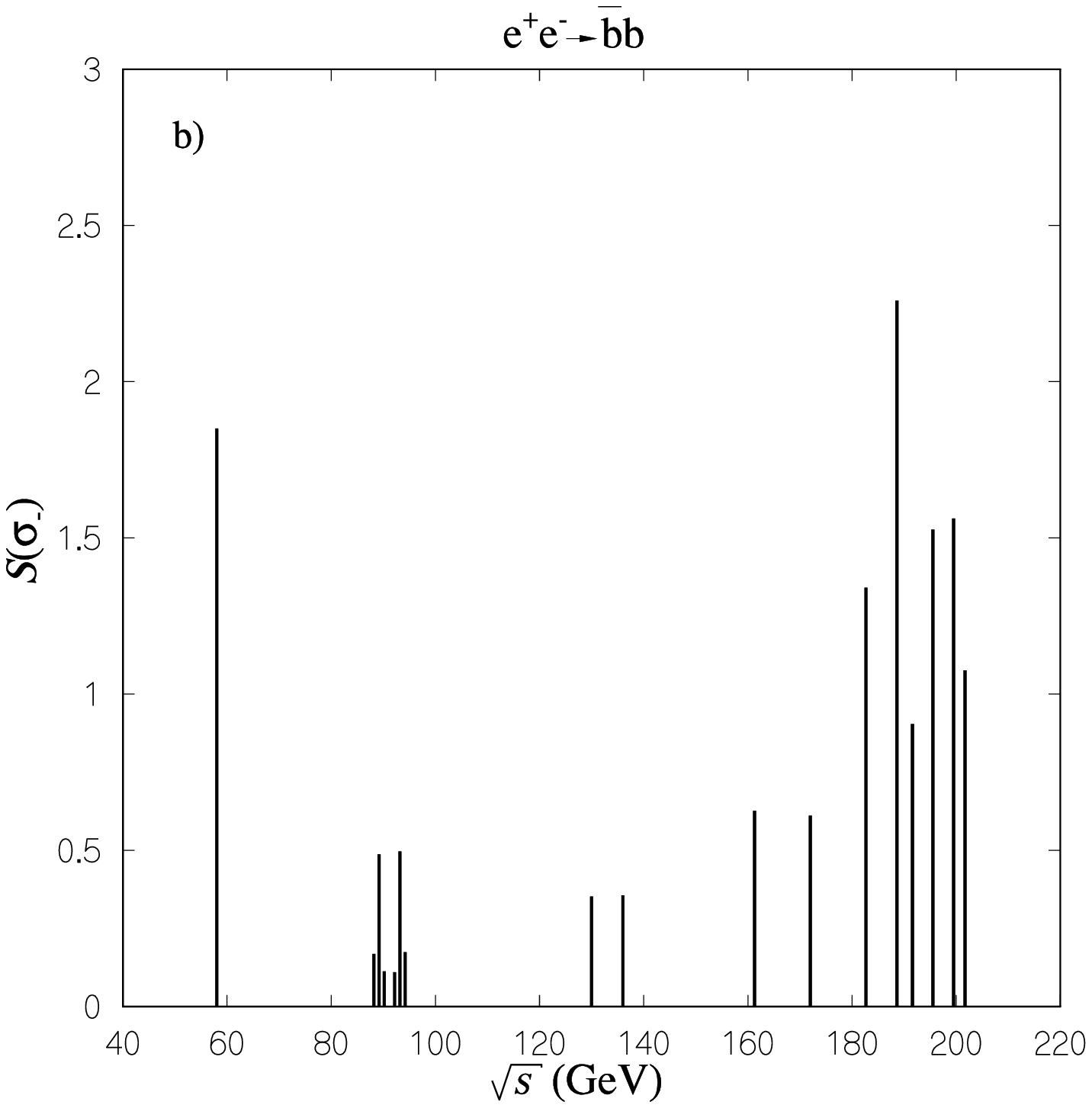}}}
\end{picture}
\caption{Same as Fig.~\ref{Fig1}, but for $e^+e^-\to\bar{b}b$.}
\end{center}
\end{figure}
\begin{figure}[htb]
\refstepcounter{figure}
\label{Fig3}
\addtocounter{figure}{-1}
\begin{center}
\setlength{\unitlength}{1cm}
\begin{picture}(12,6.5)
\put(-3.,0.0)
{\mbox{\epsfysize=9cm\epsffile{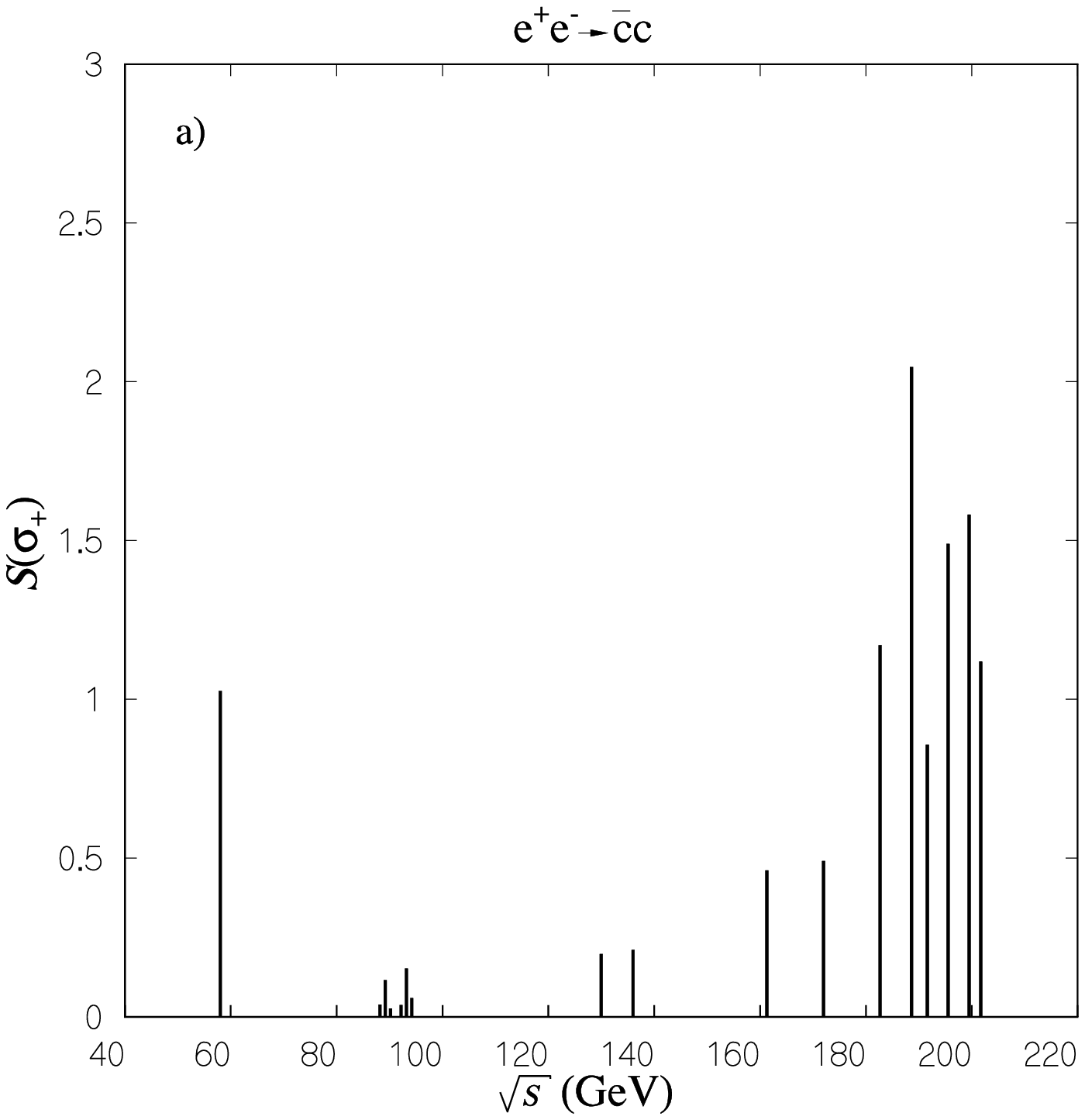}}
 \mbox{\epsfysize=9cm\epsffile{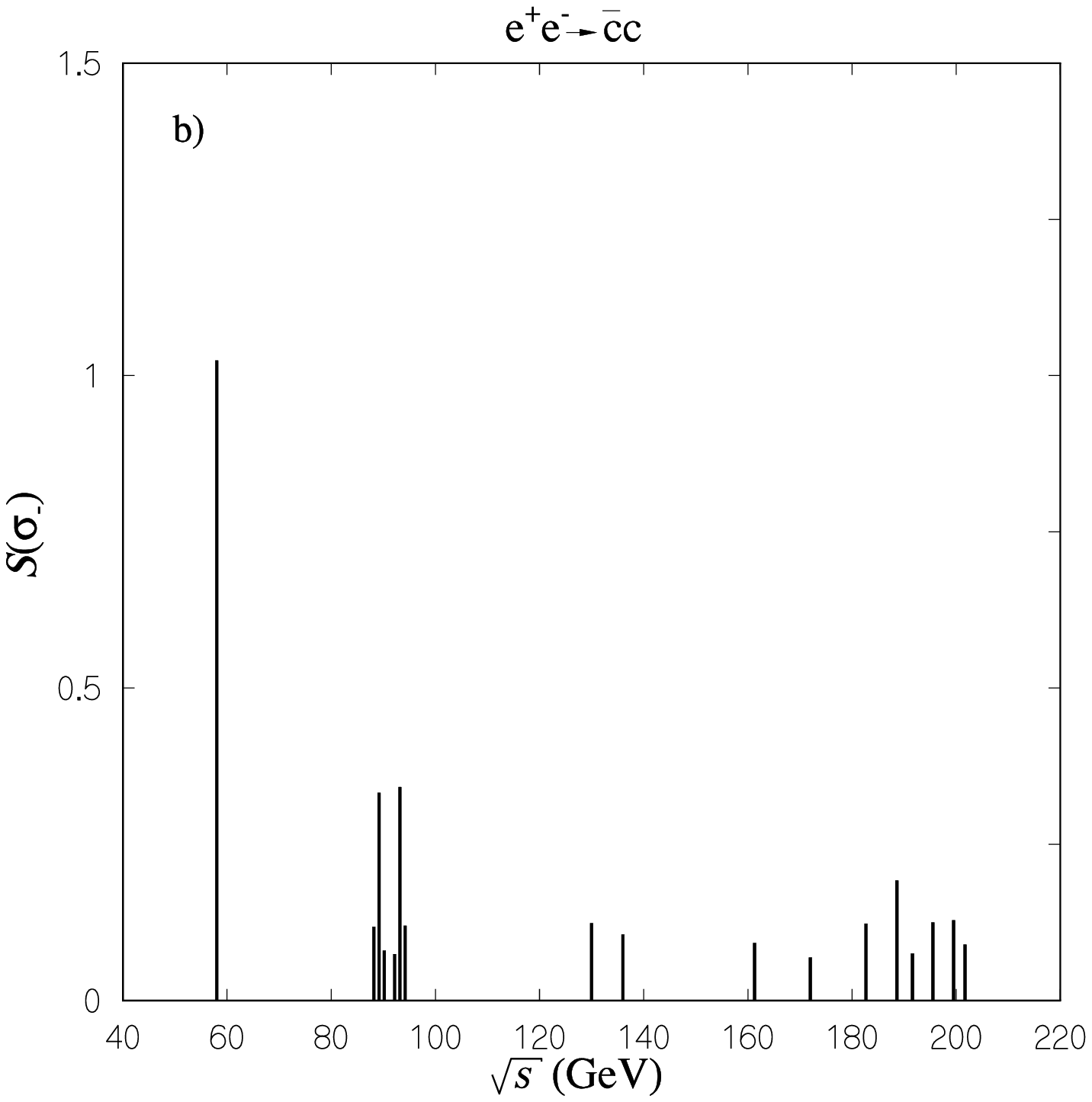}}}
\end{picture}
\caption{Same as Fig.~\ref{Fig1}, but for $e^+e^-\to\bar{c}c$.}
\end{center}
\end{figure}
Clearly, higher numerical values of $\cal S$ indicate
higher sensitivity to the relevant contact interaction parameter. One can
easily see from the previous equations that the small values of $\cal S$
around $90$~GeV reflect the zero in the real part of the $Z$ propagator at the
peak, and that, in general, $\cal S$ is small in the vicinity of a zero of SM
amplitudes. Although being just illustrative examples, Figs. 1-3 indicate that
the sensitivity significantly depends both on the reaction channel and, quite
important, on the energy. Also, they show the role of data at ``low'' energy,
where the sensitivity can be relatively high.
\par
Indeed, assuming, as it is natural, that no deviations from the SM are
observed within the experimental accuracies, constraints on the contact
interaction parameters $\epsilon_{\alpha\beta}$ can be obtained from the
inequality
\begin{equation}
\vert\Delta{\cal O}\vert < \delta{\cal O},
\label{deltacal}\end{equation}
that, using Eqs.~(\ref{s+})-(\ref{amplit}), in the planes
$(\epsilon_{\rm RR},\epsilon_{\rm LL}$) and
$(\epsilon_{\rm RL},\epsilon_{\rm LR})$ translates into the allowed areas 
enclosed by the concentric circles:
\begin{equation}
\left(\epsilon_{\alpha\beta}+a_{\alpha\beta}\right)^2
+\left(\epsilon_{\alpha^\prime\beta^\prime}+
b_{\alpha^\prime\beta^\prime}\right)^2
=\left(a_{\alpha\beta}\right)^2+\left(b_{\alpha^\prime\beta^\prime}\right)^2
\pm\kappa^2,
\label{circ}
\end{equation}
where $(\alpha\beta,\alpha^\prime\beta^\prime)=({\rm LL},{\rm RR})$ and
$({\rm LR},{\rm RL})$ for the cases of $\sigma_+$ and $\sigma_-$, respectively,
and correspondingly:
\begin{equation}
a_{\alpha\beta}=\frac{\alpha_{e.m.}}{s}A_{\alpha\beta}^{SM};\qquad
b_{\alpha^\prime\beta^\prime}=
\frac{\alpha_{e.m.}}{s}A_{\alpha^\prime\beta^\prime}^{SM};
\qquad
\kappa^2=\left(\frac{\alpha_{e.m.}}{s}\right)^2
\frac{4}{N_C\sigma_{pt}}\delta\sigma_{\pm}.
\label{circ1}\end{equation}
These relations show that both the centre and the radii of the circles
$R_1,R_2=\sqrt{a^2+b^2\pm\kappa^2}$ are determined by the SM helicity
amplitudes
and depend on energy, while the width of the allowed area is determined by the
experimental uncertainty $\delta\sigma_\pm$.\footnote{Occasionally, depending
on particular values of $s$ and $\delta\sigma_\pm$, one might have
$R_2=0$, in which case the allowed region is a circle of radius
$R_1$.} Therefore, in principle, the combination of two (or more) such allowed
regions at different energies can lead to a reduced allowed
region and, ultimately, to model-independent bounds on the contact interaction
coupling constants.
\par
In practice, to perform such combination and derive the numerical constraints
on $\epsilon_{\alpha\beta}$, we define a $\chi^2$ as 
\begin{equation}
\chi^2=\sum_i\left(\frac{\Delta{\cal O}_i}{\delta{\cal O}_i}\right)^2, 
\label{chi}
\end{equation}
where $i$ runs over the 17 data samples collected at the different energies
from $58$~GeV to $202$~GeV 
listed in Table 1, and, under the previous assumprion that no deviations are
shown by those data, we impose $\chi^2<\chi_{\rm CL}^2$, where the actual
value of $\chi_{\rm CL}^2$ specifies the desired ``confidence'' level.
As the two separate cases, $\sigma_+$ and $\sigma_-$, depend on two
independent free parameters, we choose $\chi_{\rm CL}^2=6$ as consistent with a
two-parameter analysis. The resulting bounds are depicted in Figs.~4-6.
In these figures, the dashed contours result from LEP2 data, dash-dotted ones
correspond to TRISTAN data, the full line is obtained from combination
of LEP1 and LEP2 results and, finally, the shaded area is the result of the
combination of all experiments. One can clearly see the significant role
of such combination of
measurements at different energies in deriving a restricted, model-independent, 
region allowed to the
contact interaction coublings around the SM point $\epsilon_{\alpha\beta}=0$.
\par
Indeed, the horizontal and vertical arms of the crosses in Figs.~4-6, 
intersecting at 
$\epsilon_{\alpha\beta}=0$, may be considered as qualitative indications of  
bounds obtained by taking one non-zero contact 
interaction coupling at a 
time\footnote{Actually, in this one-parameter case, one should rescale the 
value of $\chi_{\rm CL}^2$ to $\chi_{\rm CL}^2=4$.}, 
a procedure testing the individual models. In this regard, we
notice that, although on their own giving much less stringent bounds than LEP2
data in this kind of one-parameter analysis, the ``low'' energy TRISTAN data,
when combined with the former ones, play an essential role in
severely reducing the allowed area in the present model-independent
procedure taking all contact coupling constants simultaneously into account. 
\par 
Clearly, the analysis presented here in rather phenomenological, in the sense
that numerical results are not derived by a conventional fitting procedure to
$\sigma_+$ and $\sigma_-$. Nevertheless, since the
available experimental data for $\sigma$ and $A_{\rm FB}$ do not show 
deviations from the SM within the accuracies, we expect that the bounds on 
contact interaction parameters from such a fitting procedure would not 
significantly differ from those in Figs.~4-6.

\begin{figure}[htb]
\refstepcounter{figure}
\label{Fig4}
\addtocounter{figure}{-1}
\begin{center}
\setlength{\unitlength}{1cm}
\begin{picture}(12,8)
\put(-3.,0.0)
{\mbox{\epsfysize=9cm\epsffile{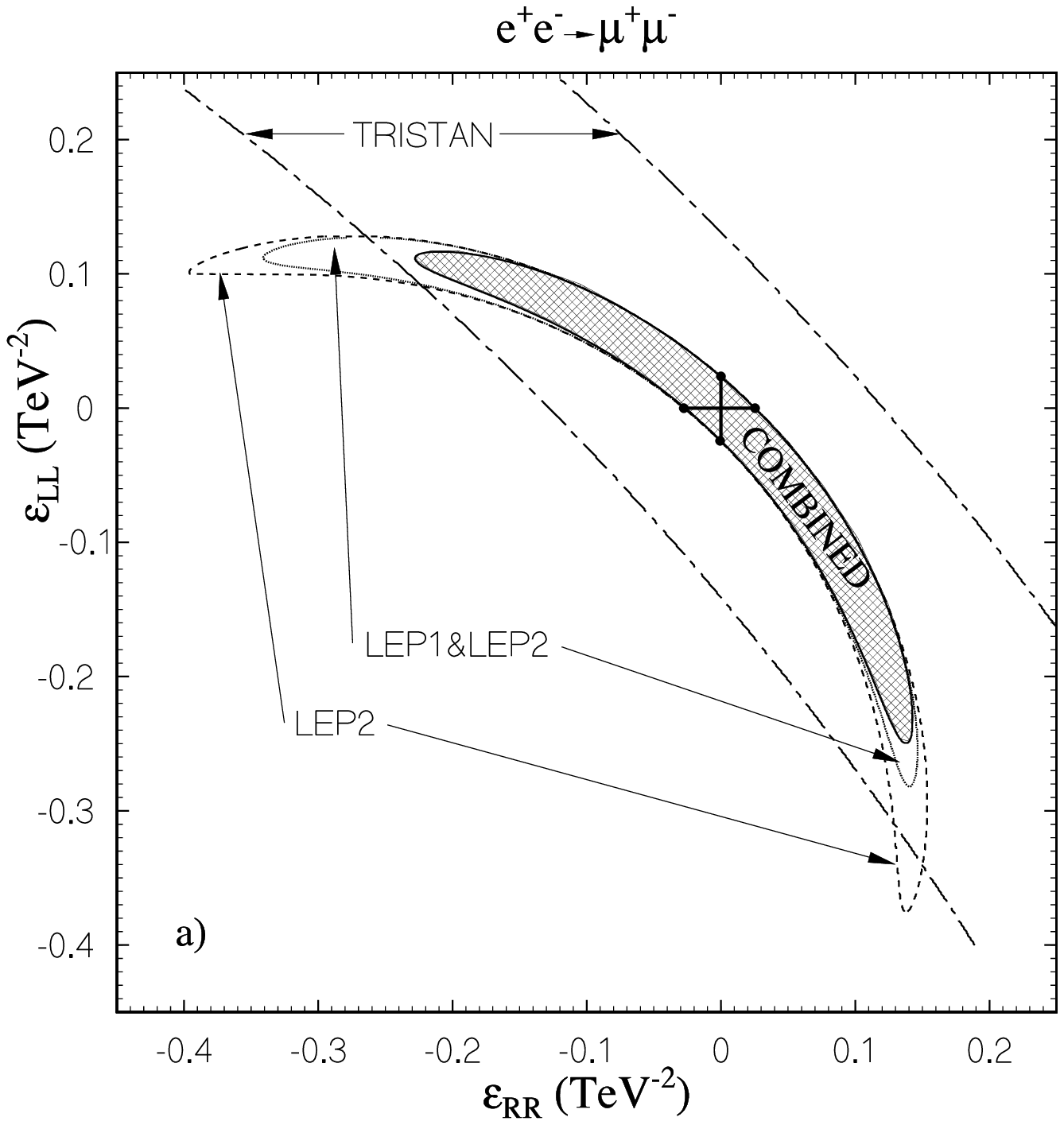}}
 \mbox{\epsfysize=9cm\epsffile{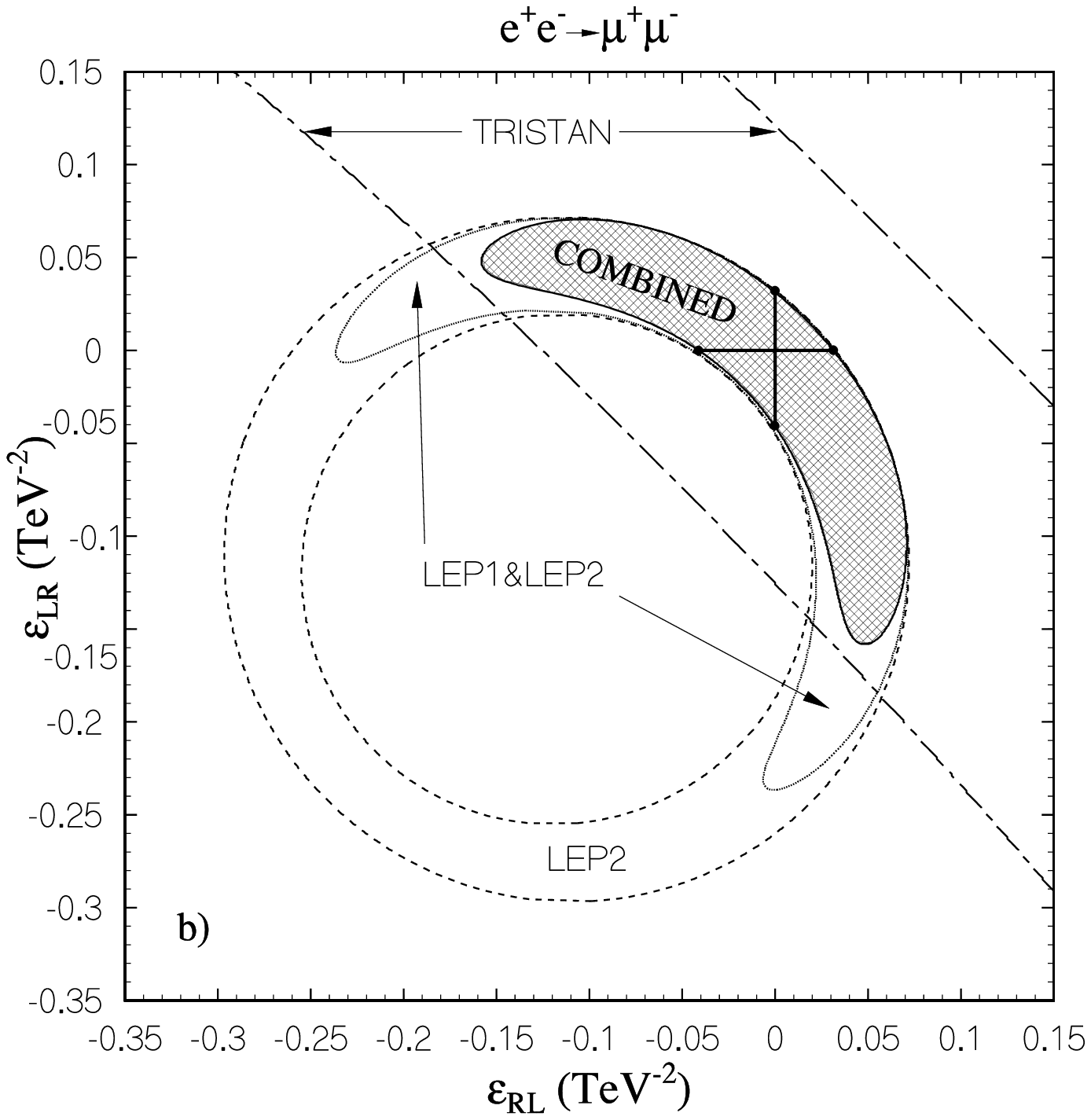}}}
\end{picture}
\vspace*{-3mm}
\caption{Allowed areas at 95\% C.L. on leptonic contact interaction parameters
in the planes ($\epsilon_{\rm RR},\epsilon_{\rm LL}$) (a) and
($\epsilon_{\rm RL},\epsilon_{\rm LR}$) (b), obtained from $\sigma_+$ and
$\sigma_-$, respectively.
}
\end{center}
\end{figure}
\begin{figure}[htb]
\refstepcounter{figure}
\label{Fig5}
\addtocounter{figure}{-1}
\begin{center}
\setlength{\unitlength}{1cm}
\begin{picture}(12,8)
\put(-3.,0.0)
{\mbox{\epsfysize=9cm\epsffile{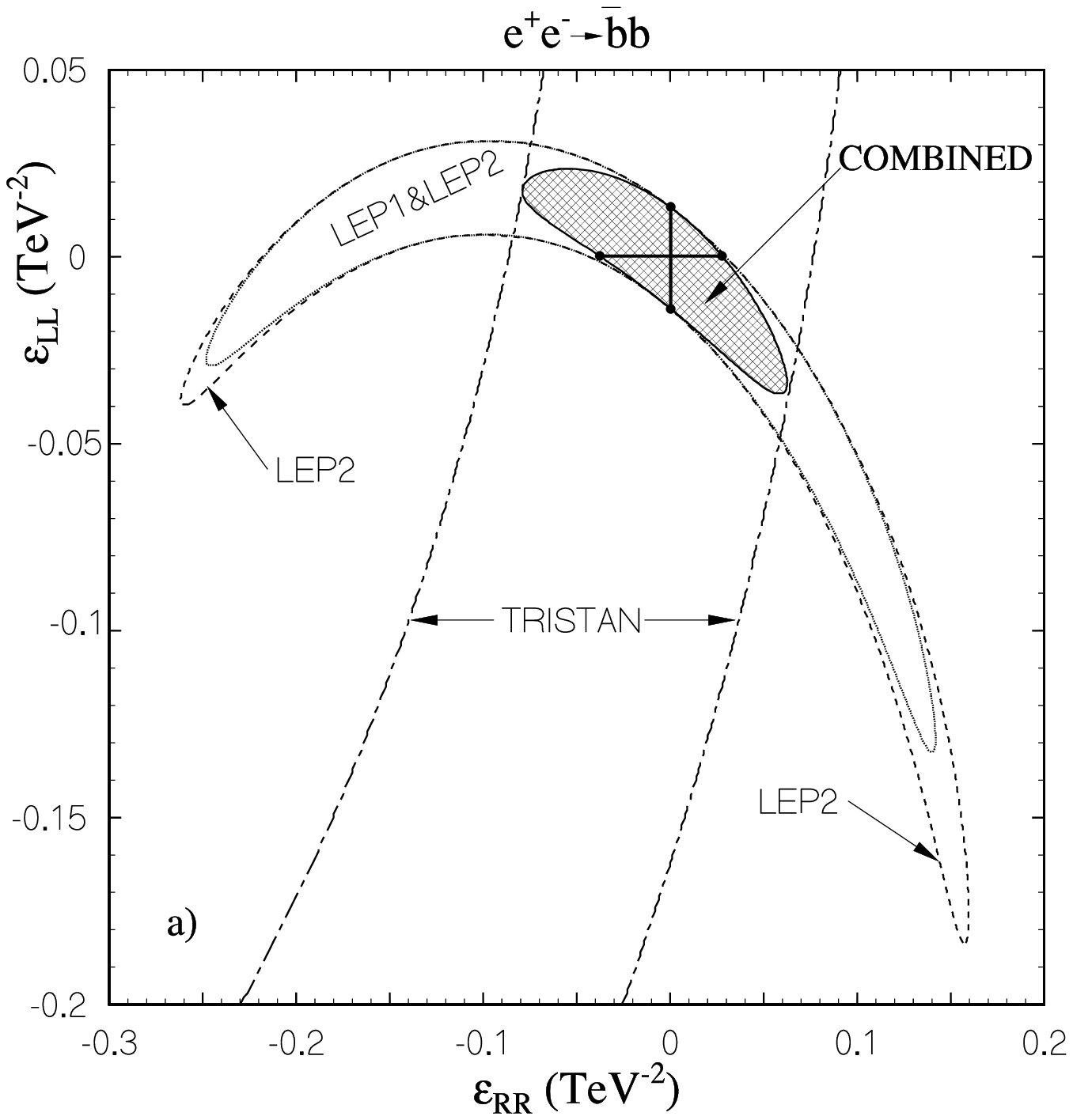}}
 \mbox{\epsfysize=9cm\epsffile{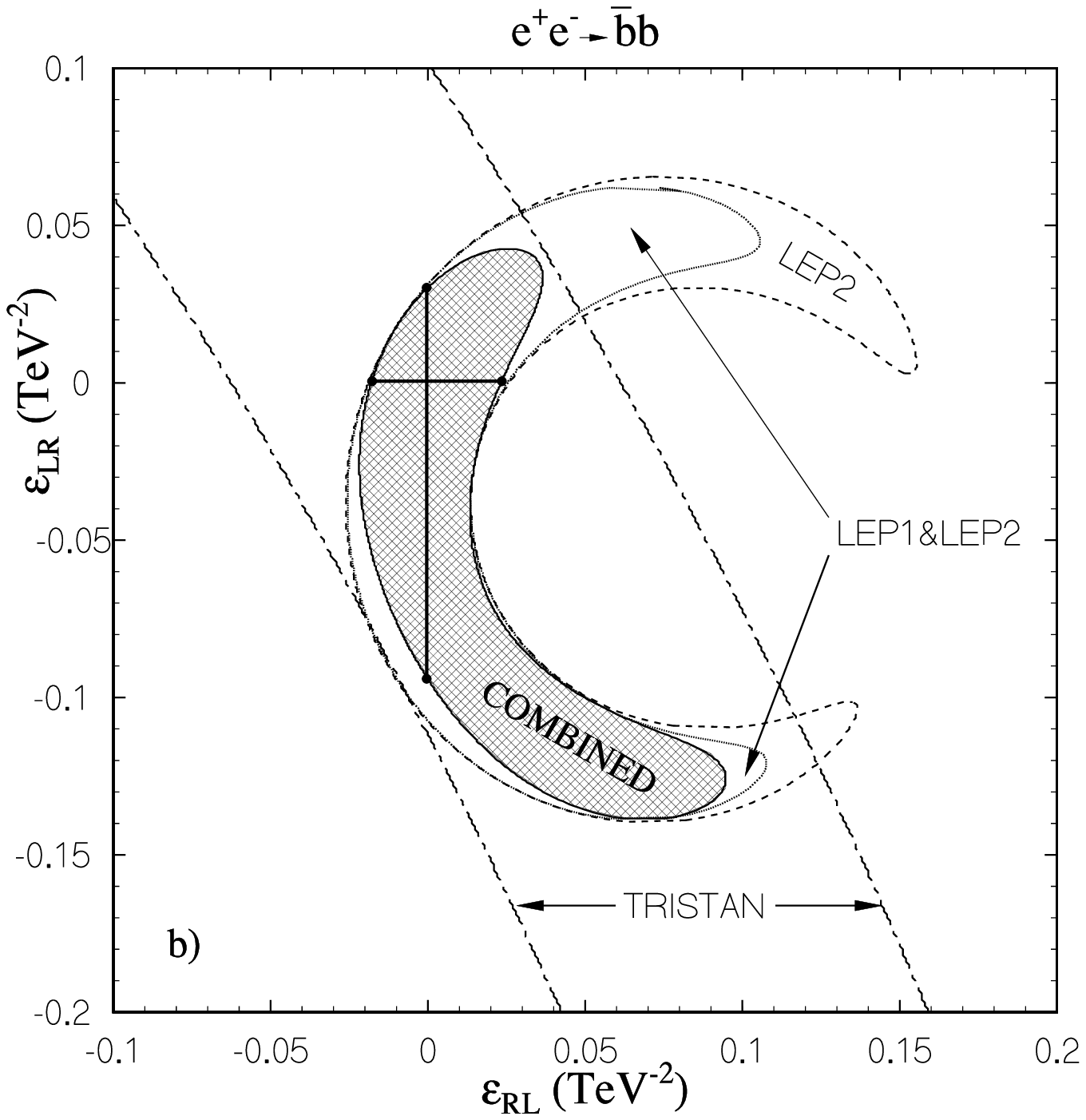}}}
\end{picture}
\vspace*{-3mm}
\caption{
Same as Fig.~\ref{Fig4}, but for $e^+e^-\to\bar{b}b$.
}
\end{center}
\end{figure}
\begin{figure}[htb]
\refstepcounter{figure}
\label{Fig6}
\addtocounter{figure}{-1}
\begin{center}
\setlength{\unitlength}{1cm}
\begin{picture}(12,8)
\put(-3.,0.0)
{\mbox{\epsfysize=9cm\epsffile{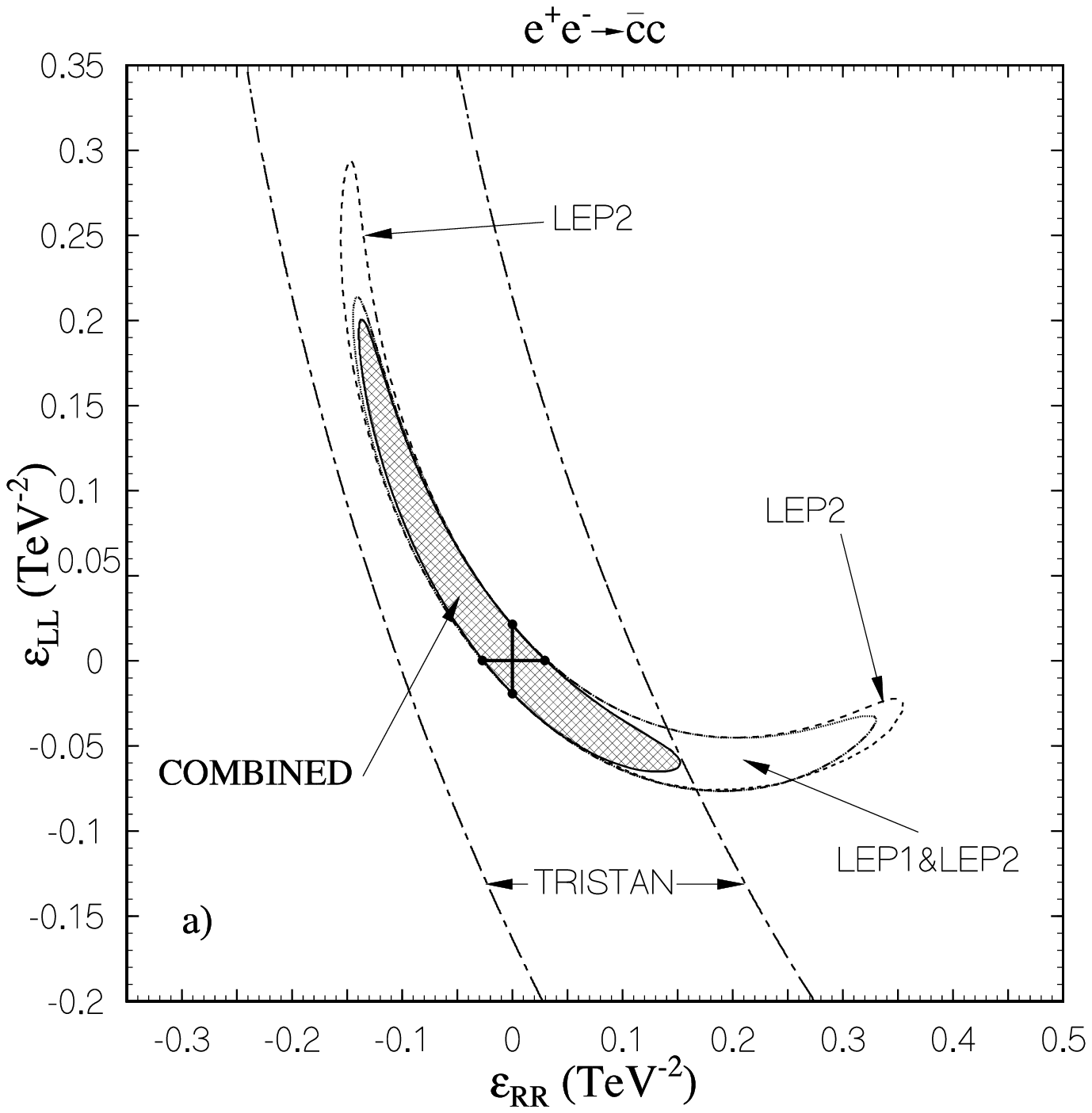}}
 \mbox{\epsfysize=9cm\epsffile{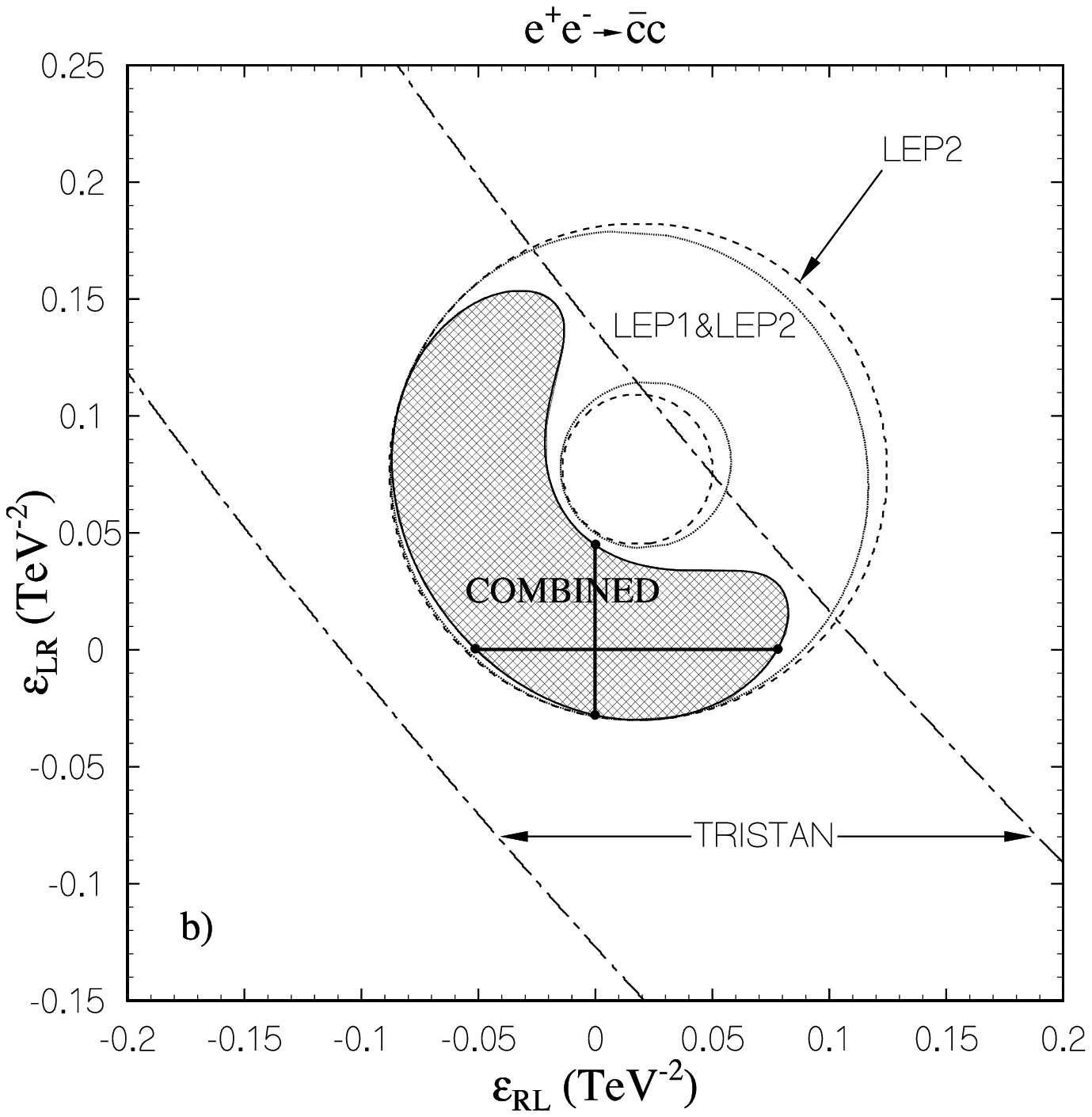}}}
\end{picture}
\vspace*{-3mm}
\caption{Same as Fig.~\ref{Fig4}, but for $e^+e^-\to\bar{c}c$.}
\end{center}
\end{figure}



\newpage
\clearpage

\begin{thebibliography}{99}

\bibitem{lep} See, e.g.:
D. Abbaneo et al., The LEP Collaborations, CERN-EP-2000-016.

\bibitem{eichten}
E. J. Eichten, K. D. Lane, M. E. Peskin,
Phys.\ Rev.\ Lett.\ {\bf 50} (1983) 811; \\
R. R\"uckl, Phys.\ Lett.\ {\bf B 129} (1983) 363.

\bibitem{barger} V. Barger, K. Cheung, K. Hagiwara,
D. Zeppenfeld, Phys. Rev. D {\bf 57} (1998) 391; \\
D. Zeppenfeld, K. Cheung, preprint MADPH-98-1081, hep-ph/9810277.

\bibitem{altarelli}
G. Altarelli, J. Ellis, G. F. Giudice, S. Lola, M. L. Mangano,
Nucl.\ Phys.\ {\bf B 506} (1997) 3; \\
R. Casalbuoni, S. De Curtis, D. Dominici, R. Gatto, Phys.
Lett. {\bf B 460} (1999) 135; \\
V. Barger, K. Cheung, Phys.\ Lett.\ {\bf B 480} (2000) 149.

\bibitem{kroha} H. Kroha, Phys. Rev. D {\bf 46} (1992) 58.

\bibitem{pankov} A.A. Babich, P. Osland, A.A. Pankov, N. Paver,
Phys.\ Lett.\ {\bf B 476} (2000) 95; {\bf B 481} (2000) 263.

\bibitem{zeppenfeld2}
B. Schrempp, F. Schrempp, N. Wermes,
D. Zeppenfeld, Nucl. Phys. {\bf B 296} (1988) 1.

\bibitem{hollik}
M. Consoli, W. Hollik, F. Jegerlehner, {\it in} Z physics at LEP1,
G. Altarelli, R. Kleiss, C. Verzegnassi (Eds.), vol.1, p.7, 1989.

\bibitem{altarelli2}
G. Altarelli, R. Casalbuoni, D. Dominici, F. Feruglio, R. Gatto,
Nucl.\ Phys.\ {\bf B 342} (1990) 15.

\bibitem{zfitter}
S. Riemann, FORTRAN program ZEFIT Version 4.2; \\
D. Bardin et al., preprint DESY 99-070, hep-ph/9908433.

\bibitem{tristan}
M. Miura et al., Phys. Rev. D {\bf 57} (1998) 5345; \\
T. Arima et al., Phys. Rev. D {\bf 55} (1997) 19; \\
K. Ueno et al., Phys.\ Lett.\ {\bf B 381} (1996) 365;\\
D. Stuart et al., Phys. Rev. D {\bf 49} (1994) 3098; \\
K. Abe et al., Phys.\ Lett.\ {\bf B 313} (1993) 288.

\bibitem{lep1}
P. Abreu et al., Eur. Phys. J. {\bf C10} (1999) 219;\\
P. Abreu et al., preprint DELPHI 98-113 CONF 175;\\
D. Abbaneo et al., The LEP/SLD Heavy Flavour Working Group,
preprint DELPHI 98-54 CONF 246.

\bibitem{lep2}
P. Abreu et al., preprint CERN-EP/2000-068;\\
K. Cie\'slik et al., preprint DELPHI 2000-038 CONF 356;\\
K. Cie\'slik et al., preprint DELPHI 2000-129 CONF 428;\\
A. Behrmann et al., preprint DELPHI 2000-036 CONF 355;\\
P. Abreu et al., Eur. Phys. J. {\bf C11} (1999) 383;\\
A. Behrmann et al., preprint DELPHI 99-58 CONF 247.

\end{thebibliography}
\end{document}